# What's the worth of a promise? Evaluating the indirect effects of a program to reduce early marriage in India

Shreya Biswas

Assistant Professor, Department of Economics

Birla Institute of Technology Pilani, Hyderabad Campus

(Email: shreya@hyderabad.bits-pilani.ac.in)

Upasak Das

Presidential Fellow in Economics of Poverty Reduction

Global Development Institute, University of Manchester

And

Research Affiliate

Centre for Social Norms and Behavioral Dynamics

University of Pennsylvania

(Email: upasak.das@manchester.ac.uk)

(April 25, 2021)

What's the worth of a promise? Evaluating the indirect effects of a program to reduce early marriage in India

*Abstract*

*One important dimension of Conditional Cash Transfer Programs apart from conditionality is the provision of continuous frequency of payouts. On the contrary, the Apni Beti Apna Dhan program, implemented in the state of Haryana in India from 1994 to 1998 offers a "promised" amount to female beneficiaries redeemable only after attaining 18 years of age if she remains unmarried. This paper assesses the impact of this long-term financial incentivization on outcomes, not directly associated with the conditionality. Using multiple datasets in a triple difference framework, the findings reveal a significant positive impact on years of education though it does not translate into gains in labor participation. While gauging the potential channels, we did not observe higher educational effects beyond secondary education. Additionally, impact on time allocation for leisure, socialization or self-care, age of marriage beyond 18 years, age at first birth, and post-marital empowerment indicators are found to be limited. These evidence indicate failure of the program in altering the prevailing gender norms despite improvements in educational outcomes. The paper recommends a set of complementary potential policy instruments that include altering gender norms through behavioral interventions skill development and incentives to encourage female work participation.*

# 1. Introduction

Cash incentives through Conditional Cash Transfer (CCT) programs have emerged among the widely adopted interventions to improve human capital, reduce the gender gap in education and break the vicious cycle of poverty. In its most canonical form, these programs offer cash transfers which are conditional to females being unmarried till a threshold age, enrolled in schools or in some cases, also linked to attendance or academic outcomes. The income effect through the cash component along with the conditionality helps in overcoming the credit constraint and lowers the opportunity cost of female education or remaining unmarried (Fiszbein and Schady, 2009; Rubio-Codina, 2010; Baird et al., 2013; Das and Sarkhel, 2020). Empirical evidence on the effectiveness of these programs has largely been equivocal about the benefits at least in some indicators of educational outcomes or deferring child marriage (Banerjee et al. 2017; Fiszbein and Schady 2009; Glewwe and Olinto, 2004; Schultz, 2004; Maluccio and Flores, 2005; Behrman, et al., 2011; Dey and Ghoshal, 2021).

An important dimension of these programs apart from conditionality is the provision for continuous frequency of payouts which in many cases is yearly. The idea is to ensure short-term consumption smoothening with the conditionalities incentivizing longer-term human capital accumulation apart from reducing the gender gap and attaining better development outcomes for females. One of the concerns of such CCTs is the regular enforcement of conditionalities, especially in countries with weak institutions (Rinehart and McGuire, 2017). Studies have indicated that the costs of complying with periodic conditionalities can often be burdensome for the beneficiaries limiting the effectiveness of CCTs in various set-up (Brauw and Hoddinott, 2011; Heinrich and Brill, 2015). Contrary to this, other programs, instead of these continuous payouts, offer longer-term financial incentives to achieve specific objectives with fewer conditionalities that encourage higher investments in female education apart from reducing son preference and improving marital outcomes. These programs often depend on a "promise" of a lumpsum payment at the end of a fixed term with limited but related conditionality. On the ones offering promised payment to the female unmarried beneficiary after attaining a certain threshold age, policymakers expect direct benefits through a reduction in the prevalence of early marriage, Additionally, it is also expected that indirect gains can be derived through higher parental investment in human capital accumulation before payment of the earmarked amount. This paper evaluates one such program, the *Apni Beti Apna Dhan* (ABAD), which translates to "Our Daughter, Our Wealth", implemented in India by assessing the effects on human capital indicators that include educational and labor market outcomes, which are not directly linked with the conditionality. In the process of identifying the drivers of the observed effects, it also examines the program effects on several dimensions of well-being that include marital outcomes, fertility choices, time allocation for daily activities and empowerment indicators.

The ABAD program was implemented in the Indian state of Haryana in the year 1994 and spanned till 1998. Households belonging to scheduled castes (SCs) and other backward castes (OBCs) or those who are officially categorized as poor are eligible for being beneficiaries of the program.[1] Under its provisions, eligible parents from these households, who give birth to daughters, are offered an immediate financial grant along with a long-term savings bond redeemable only after the daughters turn 18 years and remain unmarried.[2] The objective of the program was to reduce son preference at birth by incentivizing female birth while also lowering the opportunity cost of early marriage through the monetary "promise" of the financial bond. In the process of this rightward shift in the age of marriage, it is expected that parents would get their daughters educated that in turn may benefit them through higher participation in the labor market and changing gender norms leading to better bargaining power and empowerment post marriage.

There are multiple channels through which educational effects of such programs that do not explicitly carry any conditionality of school enrollment or attendance can manifest. Firstly, because of the added monetary incentivization and "promise", it is expected that the parents would "value" their daughter more, leading to higher private investments in human capital. However, such a premise would not hold if parents hold misguided beliefs about the process of investing in the daughter's education and subsequent returns from these investments. For instance, if parents' perceived education elasticity of long-term earnings is low; it may prevent them from sending daughters to school without the conditionality. Further, with weaker institutions, the net monetary benefit that the parents expect to receive can be much lower than the gross promised amount because of the costs that one may have to incur while redeeming the money. Secondly, it is also possible that the parents may discount the future at differential rates for sons and daughters in the family; specifically they may have systematically higher discount rates for girls. The intrinsic value of daughter's education may be low from the point of view of parents and low levels of educational investment for daughters may stand out as rational choice. The sons co-residing with parents are more likely to take care of them and provide informal old-age insurance whereas returns to girl's education will be enjoyed by her husband's family after marriage. Conditionality can then incentivize higher schooling for daughters through which the socially inefficient outcomes in terms of gender disparity in human capital investment can be addressed. Thirdly, evidence indicates equivocally that household work can stand as a significant deterrent to girl's education (Levinson and Moe, 1998). Without an associated conditionality on education, the shadow wage of these activities may not reduce which may remain as a deterrent for parents to send their daughters to schools (Bourguignon et al. 2003; Garcia and Saavedra 2017; Parker and Todd, 2017).

Despite these potential reasons, which indicate that there may not be sufficiently large gains in education because of ABAD, the conditionality of the lumpsum transfer on being unmarried until 18 years may in other ways also encourage higher schooling for the daughters. In a state like Haryana with entrenched patrilineal norms where early marriage is socially acceptable, increasing private investments in human capital for the daughters may offset the cost of delayed marriage because of the conditionality. From this perspective, the program can potentially yield higher gains on education, even though the overall program effects are unclear and this warrants an empirical investigation. Additionally, whether positive education gains (if any) translate into a higher likelihood of female work participation also remains ambiguous. Drawing from Human Capital theory, higher levels of education may lead to higher wages in the labor market beyond the reservation wage encouraging more women into paid work. However, if this relationship is non-linear in nature, it would only be above a threshold level of educational attainment that the returns to wages would start manifesting. On the contrary, if higher educational attainment leads to improved relative bargaining power in the household, women may prefer leisure or home production over going outside and working. Moreover, the marginal non-market returns to education for a woman might be higher than the market returns. The prospect of the gains in female education leading to better marital outcomes in terms of better-educated husbands or wealthy family may induce parents to get her married early. If attitudes towards women as traditional caregivers remain unaltered despite the interventions, it is likely that the gains in education may not get translated into higher participation in labor market.

This paper uses data from multiple representative sources to estimate the direct and indirect effects of the long-term "promise" of the monetary incentive provisioned under the ABAD program. Having observed a significant reduction in child marriage because of the program, first it examines the program effects on the level of educational attainments for the beneficiaries and additionally tests if the gains translate into higher labor market participation. Further, it assesses the potentials reasons and channels that explain the observed program effects. In particular, we use nationally and state representative household survey data from the fourth wave of the National Family and Health Survey (NFHS-4) conducted in 2015-16 to assess the education effects. Apart from this, we employ the two rounds of the Periodic Labor Force Survey (PLFS) conducted in 2017-18 and 2018-19 and the Time-Use Survey (TUS) conducted in 2019 to gauge the labor market effects.

To tease out the unbiased treatment effects, we exploit the exogenous timing of the ABAD implementation (1994 to 1998) in Haryana and the particular age cohort of the females who were born in this period along with other eligibility criteria. For each of the survey data used, we identify the age cohort of the females (depending on the survey year), who might be potential beneficiaries of the program. Females from the elder age cohort are taken in the control group. The neighboring state of Punjab is taken as a further control unit for Haryana. A triple difference regression strategy is employed to estimate the average intent-to-treat effects for the female beneficiaries with the elder non-beneficiaries in Haryana relative to the males and then net out the similar changes in Punjab. This estimation strategy would account for two possible factors that may confound the causal estimate: changes in the educational levels for females relative to the males of the same age cohort residing in Haryana and then changes in the outcome variable for the females across the two states. Importantly, we argue this strategy can control for a generic systematic push for attaining more education within Haryana over time and also across state interventions that can differentially affect the females relative to the males. To estimate the treatment effects on the labor market, we use the PLFS and TUS survey; hence, we are able to assess the gains (if any) not only on work participation but also time allocated for paid work outside home.

The findings indicate a significant and positive effect on years of education while having no bearing on labor participation for the beneficiaries. Importantly, we find no systematic change in pre-program trends between the treatment and control units through parallel trend assumption tests that allow us to look at the regression estimates from a causal lens. Further, we conduct a number of robustness and falsification checks that ensure our estimates are robust and indeed give treatment effects. Interestingly, we are largely able to replicate our results across rural and urban sectors or different caste groups and wealth quintiles, indicating that the effects have been mostly uniform. To identify the potential reasons for observing no impact on work participation despite a significant increase in the levels of education, we consider three channels that are indicative of no alteration in gender norms because of the program. First, the likelihood of gaining higher secondary education because of the ABAD is found to be limited, though an increase in the likelihood of completing secondary education is observed. If the relationship of work participation with education is non-linear, limited effect on acquiring higher levels of education may pose as a significant deterrent of getting work for the beneficiaries. Second, using the TUS, we find no evidence of the potential beneficiaries being more likely to substitute their labor work with leisure, socializing or self-care. Third, we consider age at marriage and fertility age to explain the insignificant bearing of higher educational levels evident because of ABAD on work participation. While the likelihood of child marriage has significantly reduced owing to the intervention, which happens to be directly linked to the program, chances of marriage at 18 or 19 does not decrease and there is some evidence of higher likelihood of marrying on turning 19 years. Also, the effect on age at the first birth of these beneficiaries is statistically indistinguishable from zero. In addition, we observe no changes in the indicators pertaining to empowerment or gender attitude because of ABAD. These set of evidence indicates limited impact of the intervention in terms of altering the social norms especially those related to gender and their traditional role as caregivers and to a lesser extent as earners.

The paper contributes to literature and policy in multiple ways. First, this, to our knowledge, is among the first rigorous evaluation of the ABAD intervention that examines potential beneficiaries from the entire state of Haryana and also those who have potentially benefitted from the program. In general, our study contributes to the limited literature on long-term effect of CCTs (Millan et al., 2020, Peruffo and Ferreira, 2017; Baez and Camacho, 2011). Second, we add to the evidence on the indirect effects of CCTs (Hasan, 2016) wherein we study the economic implications of how parents adjust the educational investments for their daughters in the presence of a program that promises a financial payment and conditionality that are not directly linked with education. Importantly, we document these evidences in the context of a setting that has highly entrenched patriarchal norms. Third, it adds to the literature that documents evidence on demand-based interventions to improve female education and their well-being that are relevant in the context of the Global South. The findings of the study enable us to comment on the policy implications, especially those on the provisions and conditionality of transfers related directly or indirectly to human capital accumulation.

The structure of the paper is as follows. Section 2 discusses the ABAD program and the context of Haryana relative to other states. Section 3 and section 4 present the description of the data and the empirical strategy used in the paper respectively. Section 5 discusses the results from the regression along with the set of robustness checks and falsifications and heterogeneous effects. Section 6 discusses the potential channels that may explain the observations. The paper concludes with discussions on policy implications through section 7.

## 2. The program and the context

Haryana is a northern state of India that has a high per-capita income but with one of the lowest child sex ratio in the country. Data from Census 2011 indicates that the child sex ratio for children in the age group of 0-6 years, in Haryana was 830 girls per 1000 boys against the national average of 914 girls[3]. Of note is the fact that this did not show a substantial change when compared with the Census 1991 and 2001 figures (820 and 879 girls for every 1000 boys respectively[4]). Such skewed child sex ratio is often attributed to pervasive son preference among Indian families, which appear to be more prevalent in the northern states of India (Dyson and Moore, 1983; Clark, 2000; Klaus and Tipandjan, 2015). Studies indicate strong son preference along with sex-selective abortions in northern states reduce the value of girls within families and increases the risk of mortality of girls (Arnold et al., 2002). Along with this, the prevalence of child marriage has been historically high in Haryana (Panchal et al. 2020). According to the NFHS-2 survey conducted in 1998-99, about 23% of the women in the age group 15-19 years during the survey were already married. About 27% of the women aged between 45 to 49 years were married before they attained 15 years.[5]

To address the adverse child sex ratio and reduce child marriage in the state, the government of Haryana introduced a state sponsored conditional cash transfer (CCT) scheme called the *ABAD* in 1994. The program eligibility conditions required that the households are those domiciled in Haryana and belonging to the deprived social groups, including the SCs and OBCs and all others lying below poverty line (BPL). Upon the birth of a daughter, the family would be entitled to receive Rs.500 (around $15.5 in 1994[6]) along with a Unit Trust of India certificate (bond) with a maturity value of Rs. 25,000 (about $380[7]) redeemable only when the daughter turn 18 years and still remains unmarried. The girl gets a bonus of Rs. 5000 (about $76) if she completes primary schooling and an additional amount of Rs. 1000 ($15) if she completes eight standard schooling. The scheme was expanded in 1995 by allowing beneficiaries to receive Rs. 30,000 or Rs, 35,000 in case the maturity is deferred by two years or four years respectively.

The ABAD is different from other well-known CCTs like the *Opportunidades* in Mexico, *Bolsa Familia* in Brazil, *Juntos* in Peru, *Familias en Accion* in Colombia and school assistance program in Bangladesh among others in two aspects. First, unlike the other CCTs, ABAD did not have any conditionality of nudging families to increase investment in health and human capital of daughters even though there was a small incentive in terms of bonus to educate daughters ($76 for attaining primary education and $13 for completing education up to eight standard). Second, a unique feature is the long-term nature of the program benefits, which resembles a "promise" of monetary transfer instead of a continuous payment. Important to note is the fact that similar to zero coupon bonds, the beneficiaries of ABAD did not receive any intermediate payments till the age of 18 except for the payment of Rs.500 received just after the birth of daughters.

3. **Data and variables**

We bring in multiple datasets to evaluate the program effects on different indicators of human capital accumulations. Our main data source for assessing the effects on years of education is the fourth wave of the National Family and Health Survey (NFHS-4) conducted during 2015 and 2016 by the Ministry of Health and Family Welfare. The survey gathered information on 601,509 households across 640 districts and is one of the largest household survey in India that is representative not only at the national but also at the state and district level. The survey provides information related to age, educational attainment of individuals residing in the household along with other characteristics along with that on age at marriage, fertility outcomes as well as female autonomy indicators.

Because the program spanned between 1994 and 1998, women who were in the age bracket of 16 to 21 years and were surveyed in 2015 and women in the age bracket of 17 to 22 years who were surveyed in 2016 are identified as age-eligible women from NFHS-4. Women in the age group of 22 to 27 years and 23 to 28 years who were surveyed in 2015 and 2016 respectively form our age-ineligible control group. To identify the poor households, we use information on the Below Poverty Line (BPL) card possession. This card is often necessary to avail targeted welfare schemes that include subsidized cylinders and foodgrains among others. Nevertheless, it should be noted that this identification of poor based on ownership of BPL card is not without demerits. Studies across the country have found major inclusion and exclusion error, where a significant proportion of relatively non-poor are found to possess the card while many poorer ones do not have one (Ram et al., 2009; Drèze and Khera, 2010). However, we rely on this classification as it is largely considered an objective measure of identifying the poor and the program itself stated that the households owning a BPL card are eligible to enroll for the program and receive benefits similar to other targeted programs.[8] Accordingly, women from Haryana in the age-eligible group and either possessing a BPL card or belonging from the SC or OBC social group constitute our treated group.

To examine the program effects on labor market participation, we make use of household data from two rounds of the PLFS conducted in 2017-18 and 2018-19 by the National Sample

Survey Organization (NSSO) and the Ministry of Statistics and Program Implementation of the Government of India. For the purpose of our analysis, we pool both rounds of the cross-sectional household data of PLFS and explore the details on the employment of the individuals. The first round of the survey covered 102,113 households and the second one covered 101,579 households. To assess the labor market effects, we also employ the TUS conducted by the same organizations in 2019, which provides information about the time duration for which different activities are performed by the sampled population on the day prior to the survey. It collected information from 518,751 individuals belonging to 138,805 households. These surveys are micro-unit recorded and representative nationally as well as at the state level and can be useful in gauging the labor market scenario of the country.

The TUS data helps in both corroborating the results obtained from PLFS along with providing additional insights into female's work participation. Studies suggest that women's work is better captured by TUS, especially from developing economies in comparison to labor force surveys (Hirway and Jose, 2011; Eswaran et al., 2013). Such surveys allow us to capture multiple jobs performed by women during the day such as feeding farm animals, working as an informal worker on a nearby farm and working in own account enterprise and some of these activities can also be performed simultaneously. On the contrary, the labor force surveys can only account for the primary and secondary activity status which may not be able to capture these nuances leading to underestimation of female's work participation.

As in the earlier case, because the survey was conducted in 2017-18 and 2019 and the program spanned from 1994 to 1998, the age cohort of individuals eligible to be in the treated group are 18-23 years, 19-24 years and 20-25 years for those surveyed in 2017, 2018 and 2019 respectively (age-eligible). To identify the individuals for the control group, we use the elder cohort with the same age spacing as above. Hence individuals in the age cohort 24-29, 25-30 and 26-31 surveyed in years, 2017, 2018 and 2019 respectively, are included in the control group (age-ineligible). Because no information on BPL card possession is given in the survey, we depend on Month Per-Capita Household Consumption Expenditure (MPCE) and use cut-offs of 25% to identify households are grouped as poor.[9,10] Accordingly, females from Haryana in the age cohort as mentioned above either from the poor households as defined or from the SC or OBC community are the potential beneficiaries of the ABAD intervention. For the analysis with the TUS survey, which was conducted in 2019, females in the age cohort 20-25 years (age-eligible) from poor households as defined above or from the SC or OBC community constitute the eligible group. Females in the age cohort 26-31 years constitute the age-ineligible group.

Our main outcome variable here is the number of years of education completed and work participation. The former comes directly from the NFHS-4 dataset while the latter is derived from the PLFS and TUS. From PLFS, we use information under the "usual principal activity status" (UPS). The UPS collects data about the activity status on which the members of the sampled households spent major time during the last 365 days preceding the date of the survey. From the UPS, we grouped those members who have participated in the labor market and coded them as "1" and others as "0". In a second specification, we examined the program effects on working as skilled labor and hence recoded those individuals who have reported having worked as casual labor as their UPS as 0. The extensive information on household and individual level socio-economic and demographic characteristics allows us to use them as control variables in the regression. This includes household size, religion, wealth score/ per capita consumption expenditure along with survey time, quarter of the year, round and district dummies.[11] From the TUS that uses the International Classification of Activities for Time-Use Statistics (2016), we calculate the share of time allocated to "employment and related activities" and "production of goods for final use" on the day prior to the survey and use it as our dependent variable to capture labor market participation.[12] The basic summary statistics of treated and control group females and males from the NFHS-4, PLFS and TUS datasets are provided in Table 1.

[Table 1 here]

## 4. Empirical Strategy

We use a triple difference (DDD) regression to obtain the intent-to-treat estimate of the ABAD program on educational attainment in terms of years of education. The first difference compares the outcome based on age eligibility of females. To account for overtime state specific interventions within Haryana that may systematically increase educational attainment of the beneficiaries, we include males from the eligible households in the state in the same age-cohort.[13] In addition, to ensure that our estimates are not confounded by omitted variables that differently affect the trend in male and female education in Haryana, we use a DDD regression framework.[14] Here we include poor males and females in the age group considered residing in the neighboring state of Punjab as a control to Haryana. Punjab classifies as a relevant control group for Haryana for several reasons. First, it was carved out of the state of Punjab as a result of the Punjab Reorganization Act in 1966, implying residents of these two states share similar historical experiences.[15] Second, the city of Chandigarh is the joint current capital of both Haryana and Punjab. In addition, the literacy rate in Punjab and Haryana is not very different along with the sex ratio in these two states, which is comparable too.[16] Moreover, the crop intensity of Punjab and Haryana between the years, 2012 to 2015 is also found to be similar, all of which justify the selection of Punjab as a control state[17]. Accordingly, using Punjab as a control state to Haryana, we use a triple regression to assess the program effects as our main model for evaluation.[18] The identifying equation from our triple-difference intent-to-treat estimate is based on the following estimation model:

$$y_{ihds} = \beta_0 + \beta_1 Haryana_{ih} * A_{ihds} * Female_{ihds} + \beta_2 Haryana_{ih} * A_{ihds} + \beta_3 Haryana_{ih} * Female_{ihds} + \beta_4 A_{ihds} * Female_{ihds} + \beta_5 Haryana_{ih} + \beta_6 A_{ihds} + \beta_7 Female_{ihds} + \sum \gamma_k X_{ihsk} + \delta_d + \varepsilon_{ihs} \tag{1}$$

Where $y_{ihds}$ is the outcome variable of interest for individual $i$ from household $h$ and district, $d$ and state, $s$ (Haryana and Punjab). $A_{ihds}$ is a dummy variable that equals to one for age-eligible individuals and zero for age non-eligible men and women in the sample. The variable, $Haryana_{ih}$ which takes the value of 1 if individual $i$ from household $h$ and district, $d$ resides in the state of Haryana and 0 if he/she resides in Punjab. $Female_{ihds}$ dummy assumes the value of one for the females and zero otherwise. $X_{ihdk}$ is the set of individual and household level confounders with $k$ being the number of such confounders. Additionally, in our model we control for district fixed effects to control quality of local administration and implementation of other welfare programs. The errors are clustered at the village/ town level which is the primary sampling unit (PSU). $\beta_1$ gives the average intent-to-treat estimate and captures the causal effect of the program on the outcomes for treated females.

We test for the parallel trends assumption in our DDD model for those elder to the age-eligible individuals using the following equation:

$$y_{ihds} = \beta_0 + \beta_1 Haryana_{ih} * Age_{ihds} * Female_{ihds} + \beta_2 Haryana_{ih} * Age_{ihds} + \beta_3 Haryana_{ih} * Female_{ihds} + \beta_4 Age_{ihds} * Female_{ihds} + \beta_5 Haryana_{ih} + \beta_6 Age_{ihds} + \beta_7 Female_{ihds} + \sum \gamma_k X_{ihsk} + \delta_d + \varepsilon_{ihs} \tag{2}$$

As one may observe, this is similar to equation (1), the only difference being the variable, $Age_{ihds}$ which takes the age of the individual $i$ from household $h$ residing in district, $d$ of state, $s$. The parameter estimate of $\beta_1$ if statistically insignificant ensures non-rejection of parallel trends.

While estimating the impact on labor market participation, we apply triple difference regression as given in equation (1) and test parallel trends assumption by estimating equation (2). For the analysis with PLFS, the outcome variable, $y_{ihds}$ takes the value of 1 if the individual $i$ from household $h$ residing in district, $d$ of state, $s$ goes outside for work/ skilled work and 0 otherwise. For that with the TUS, the outcome variable is the share of time allocated to "employment and related activities" or "production of goods for own final use" the day prior to the survey for the sampled individual.

To estimate the effects on time allocation, marital outcomes or empowerment measures, we run double difference/ difference-in-difference (DD) regressions for the female sample from Haryana and Punjab with those from the age-eligible cohort as well as the age non-eligible cohort. This is because these indicators are not relevant to the men in the context of India. The regression uses the following equation to estimate the effects on time allocation in different broad activities:

$$Y_{ihds} = \beta_0 + \beta_1 Haryana_{ih} * A_{ihds} + \beta_5 Haryana_{ih} + \beta_6 A_{ihds} + \sum \gamma_k X_{ihsk} + \delta_d + \varepsilon_{ihs} \quad (3)$$

Here the notations remain similar to those in equation (1) and the regression is only run for the female samples from Haryana and Punjab. Here, the parallel trend assumption is tested through the following equation, which is similar to (2) but in a DD framework:

$$Y_{ihds} = \beta_0 + \beta_1 Haryana_{ih} * Age_{ihds} + \beta_5 Haryana_{ih} + \beta_6 Age_{ihds} + \sum \gamma_k X_{ihsk} + \delta_d + \varepsilon_{ihs} \quad (4)$$

As indicated earlier, this regression is run for the sample of females, who are elder to the age-eligible ones.

## 5. Results

As discussed, this paper evaluates the ABAD program by examining the indirect effects of this "promise" of cash transfer in terms of female education and then their participation in the labor market. Here we postulate that the gains in years of education and then the subsequent labor market effects would be contingent on whether the program has been successful in arresting child marriage. We observe from NFHS-4 data that the proportion of child marriage among treated females in Haryana fell by almost 16 percentage points compared to control group females; however, the fall in the share of child brides was considerably lesser at 7 percentage points for similar age-group in Punjab. This fall in in child marriage for ABAD beneficiaries can potentially be a direct consequence of the conditionality attached with the program, though we formally test this in section 6.3.

[Figure 1 here]

*5.1 Main results*

5.1.1 Effect on years of education

The density plots for educational attainment for the treated and the non-treated elder group of females from Haryana is given in Figure 1. It appears that the program beneficiaries have higher education attainment. We formally test if there is a causal relationship using a DDD regression framework as elucidated in equation (1). Table 2 presents the results with two specifications. In the first specification, no control variables are included. However, in the second, full set of control variables are incorporated along with the district fixed effects that capture supply-side factors like local governance, school infrastructure and teacher quality among others. The unadjusted estimates are given in column 1 and adjusted ones in column 2, which is our preferred model for evaluation of program effects. The findings indicate an increase in education by 0.81 years on the ABAD beneficiaries because of the program, which is also statistically significant at 1% level (column 2).

[Figure 1 here]

The findings can only be considered causal if the parallel trends assumption holds through which we can assess if there had been no pre-intervention trends that would have systematically affected the treated or the control state. We test for these in the DDD specification where we include an interaction between female dummy, Haryana dummy and age variable and estimate the model for the individuals in the control group age cohort as given in equation (2). Please note that age variable is taken as a continuous one here. The findings imply that we are unable to reject the null of parallel trends in this DDD specification (column 3). Additionally, we run similar tests for other age groups (22-29 and 22-32 years) as given in table 2 (columns 4and 5). The statistically insignificant triple interaction term in all these specifications suggest that the gender difference in terms of years of education in Haryana has been invariant over time in comparison to Punjab before implementation of ABAD. This ensures that the education gains we observe can indeed be interpreted as causal effect of the intervention.

[Table 2 here]

Early evidence of ABAD beneficiary households investing in girl's education had been documented by Sinha and Yoong (2009). The study using the previous waves of NFHS finds that the beneficiaries (aged 7-11 years during the time of the survey) were more likely to continue education relative to girls in Haryana who were unexposed to the program. Our inference to some extent complements this finding in 2006 and we are able to document reasonable program effects almost two decades after program implementation. This is especially pertinent because evidence indicate substantial female drop out after completion of primary education across developing countries and Haryana is not an exception here (Kingdon, 2007). Our results indicate that programs like ABAD with an objective to change the value of daughters in the family but largely devoid of any cash incentive to invest in the education of girls indeed improved the educational investments beyond the primary schooling level of about 5 years. Notably, the findings also complement Nanda et al. (2014), who analyze data from a primary survey conducted in four districts of Haryana to document that beneficiary females were likely to stay in school compared to eligible non-beneficiaries.

The qualitative effect of other variables on educational attainment is in line with what one might expect. The general trend in the increase in educational attainment over time is captured by the negative relationship with age. We further observe that individuals belonging to religious minorities (non-Hindus) have lower educational attainment. Additionally, the ones from richer households (with higher wealth score) and those residing in smaller households have higher educational attainment on average.

5.1.2 Effect of labor participation

Our initial estimate indicates a definite positive impact on educational attainment captured through years of education. Next, we test if these gains translate into better outcomes for the beneficiaries in the labor market. For this purpose, as indicated earlier, we use the PLFS survey data from 2017-18 and 2018-19 and TUS data for 2019. We use the DDD regression framework as have been used to evaluate the program effects on education. Table 3 presents the results for the three outcome indicators we outlined: participation in outside work, participation in skilled work and share of time allocated for work participation. The regression findings, estimated with the sample of individuals in the relevant age group indicate no statistically significant impact of the intervention on these three outcomes despite an overall increase in the years of education. Our findings are found to be robust even if we increase or decrease the age of the control group. As indicated earlier, in the absence of data on BPL card possession, we considered households in the state-wise lowest 25 percentile in terms of MPCE for evaluating the impact. Important to note here is that the regression results to test parallel trends assumption indicate statistically insignificant time-variant change in the treated units and have been presented in Table 3.

[Table 3 here]

*5.2 Robustness checks and falsification tests*

In this section, we discuss a battery of checks that are applied to ensure that the educational effect of ABAD, which is found to be positive and statistically significant, is correctly identified

5.2.1 Placebo tests

We start by running placebo tests through the introduction of a pseudo-treatment for females in Haryana belonging to older age cohorts. First, we introduce a placebo treatment for 22-27 year age-group and 23-28 year age-group females in Haryana (fake treatment group) who were surveyed in 2015 and 2016 respectively. Here we consider the sample of males and females in Haryana and Punjab from BPL households or belonging to socially disadvantaged group in the age cohort of 22-34 year and accordingly all males and all females from Punjab and females in the older age cohort from Haryana form the control group for this analysis. These females in the fake treatment group were unexposed to the treatment and re-estimating equation 1 should ideally yield insignificant coefficient for the triple interaction term. As expected, we find it to be statistically insignificant even at 10% level of significance (Table A2 of Appendix). Next, we consider the sample of males and females in the age-group of 24-36 and provide placebo treatment to females from Haryana in the age cohort of 24-29 years. This as well yields similar insignificant result (column 2).

5.2.2 Placebo tests with NFHS-3

Our current analysis uses the NFHS-4 dataset, which was conducted in 2015-16 allowing us to identify women from Haryana in the relevant age cohort from poor and socially disadvantaged groups as potential beneficiaries. However, the sample of women from Haryana in the same age-cohort and group collected during the NFHS-3 survey conducted in 2005-06 would have been unexposed to the intervention.[19] Specifically, we consider women who were in the age bracket of 16 to 21 years and were surveyed in 2005 and women in the age bracket of 17 to 22 years who were surveyed in 2006 as pseudo age-eligible group from NFHS-3. Hence using a similar DDD regression for a similar sample of individuals surveyed in the NFHS-3 survey, we should not observe significant gains on education for the treated group in absence of the intervention. Column 3 of Appendix table A2 which presents the results show no significant effect on this group.

5.2.3 Sample of non-poor and upper-caste individuals

Next, we consider the sample individuals belonging to households who do not have a BPL card (non-poor) and are also from upper castes in Haryana and Punjab in the relevant age-group of 16 to 28 years. These individuals were unexposed to the intervention; hence as was the expectation earlier, re-estimation of equation 1 after placebo fixing a similar placebo treatment group should ideally fetch a statistically insignificant coefficient for the triple interaction term. This is indeed found to be the case (column 4 of table A2), which lends credence to our causal inferences that we provide.

5.2.4 Alternate way of identifying the poor

Our analysis is based on cross-section data and we identify household as poor if the household owns BPL card at the time of the survey (2015-16). However, this may not correctly identify those who were eligible to receive program benefits between the years 1994-1998. It is possible that families who own a BPL card at the time of the survey may not have owned one when the ABAD was implemented. Accordingly, we check the sensitivity of our results by varying the definition of poor. The survey provides a wealth score based on the assets owned by the households. We use this asset index to identify the poor for robustness. It is worth noting that this asset index is a stock variable and can be argued to be a reasonable indicator of income or consumption expenditure that are flow variables. Given paucity of data, we rely on the asset index which though imperfect, can be considered to be a fair enough correlate of poverty.

Accordingly, we consider wealth asset score cut-offs starting the lowest $10^{th}$, $15^{th}$, $20^{th}$, $25^{th}$ and $30^{th}$ percentiles instead of BPL card possession. The regression findings are presented in figure A3 of the appendix. Overall, these findings appear to be consistent with our initial inferences to the various ways of identifying the poor in our sample. We run similar tests using MPCE percentile cut-offs to determine the program effects on labor participation (figure A4 of the Appendix). The findings from these regressions seem to indicate no significant bearing of the intervention on female work participation.

5.2.5 Bordering districts

Similar to Muralidharan and Prakash (2017), we restrict our sample to the bordering districts of Haryana and Punjab. This analysis reduces our sample size to 8,692, comprising of total 13 districts (7 districts of Haryana and 6 districts of Punjab[20]). We are able to replicate our initial inference of positive program effects on education (Column 1, table A5 of the Appendix). Next, we add the neighboring districts of Himachal Pradesh and Uttarakhand along with the neighboring districts of Punjab[21] and find that the triple interaction term is positive and significant at 1% level of significance (column 2). Column 3 reports the result of re-estimating equation (1) after adding the neighboring districts of Uttar Pradesh along with other neighboring districts from other states and the results are similar (column 3).[22] Figure 6 of the Appendix suggests that the parallel trends assumption holds for these specifications.

5.2.6 Alternate definition of age-eligibility

The intent-to-treat estimate depends upon identifying age-eligible beneficiaries using the reported age and survey year. We check the sensitivity of our results by changing the age-eligibility criteria by including all females in the age group of 16 to 22 years residing in Haryana as our age-eligible target group. This way of identifying the treated females will include intended beneficiaries irrespective of the survey year. Similarly, all females in the age-cohort of 23 to 29 years from Haryana form our age-ineligible group of women. Additionally, all men from Haryana and men and women in the age-eligible and ineligible cohorts from Punjab form our control group. The effect of ABAD on years of education is qualitatively similar using this definition of age-eligibility (Appendix figure A3). Further, given that the ABAD program was first introduced in 1994, possible program delays and institutional failures in the initial year may imply the females born in 1994 did not get the benefits or were unaware of them. Accordingly, true beneficiaries were girls born between 1995 and 1998. To ensure that our results are not driven by those born in 1994, we exclude all individuals who were 21 years in 2015 or 22 years and surveyed in 2016 from our sample. This alternate sample again yields positive and significant effect on years of education.

*5.3 Heterogeneous Effects*

5.3.1 Sector-wise effect

We examine the differential effect of ABAD based on sector given widespread disparity in rural and urban areas in terms of education (Hnatkovska et al., 2013). Figure 2a indicates that the improved education attainment of treated females is observed both in rural and urban regions with similar marginal effects.

[Figure 2 here]

5.3.2 Caste-wise effect

One of the primary objectives of the program was to improve the condition of daughters from socially disadvantageous groups that include the SCs and OBCs. Accordingly, we examine the effects of ABAD separately for individuals from these groups along with those belonging to the upper caste group. Figure 2a, which present the results suggest that the effect is positive and strongest (over a year) for SCs, followed by 0.77 years for OBCs though the effect is found to be statistically insignificant for the poor upper castes females.[23]

5.3.3 Effect across wealth quartiles

Because our sample consists of poor as well as non-poor socially disadvantaged groups, it gives us an opportunity to examine whether the effect of education heterogeneously varies with wealth.

Qualitative evidence suggests that that the level of awareness regarding the ABAD was low (Krishnan et al., 2014) and it is possible that targeted beneficiaries from the non-poor group with a generally higher level of awareness on government schemes benefited more than the others. On the contrary, it is also possible that the poorer groups with lower levels of female education would be able to take better advantage of the intervention resulting in higher effects for the poor relative to the others. To assess this, we consider quartiles of wealth score and separately examine the effect for each group. Figure 2a suggests that the effect is largely positive and significant at 5% level of significance across all the four quartiles though effect size varies marginally. Nevertheless, our results highlight the inclusive nature of the program with benefits of close to similar magnitude accruing to beneficiaries irrespective of their wealth.[24]

5.3.4 Exposure-wise effect

We also examine the differential effect of the program based on the exposure to the program. Figure 2b suggests that after controlling for potential confounders, the difference in educational attainment of intended beneficiaries and control group is found to reduce over time. The effect on education is lower for beneficiaries born just after the program initiation and effect is on average higher after two years of the program implementation. This possibly hints at the benefits of higher exposure to the intervention that among other things can improve awareness and also fix problems related to the enabling institutions. Possibly because of low levels of awareness along with social concerns, the beneficiaries often find it difficult to register their claims and get their legal entitlements. Nevertheless, with higher exposure, these complications may ease up due to which we observe higher education gains from the program though further research is required to identify the exact reasons.

## *6.* **Possible channels/ explanations**

The non-transition of educational gains into the labor market gains has important policy implications, because of which we offer some potential explanations.

6.1 *Completion of secondary and higher secondary education*

Among the existing empirical literature which has pointed out the increasing returns to education on labor market outcomes, some have documented that the relationship is non-linear. This implies that the returns to education increase non-linearly with the labor outcomes meaning that the marginal return is higher at higher levels of education. One of the reasons that explain lower female labor participation in India is the lack of higher education attainment among them (Mehrotra and Parida, 2017). Drawing from this, we examine if the program has resulted in a higher probability for the female beneficiaries to complete secondary and higher secondary education. This also becomes important from the point of view of the provisions of the program ties financial incentive to completion of primary and $8^{th}$ standard education.

To explore this, we estimate a probit regression model in the DDD framework, the equation of which is similar to the one elucidated in equation (2). The outcome variables in the two cases are completion of secondary education or completion of higher secondary education. Accordingly, for the former, it takes the value of 1 if the sampled individual has completed secondary education and 0 otherwise; and for the latter, it takes the value of 1 if higher secondary education is completed and 0 otherwise. Column 1 of Table 4 presents the average marginal effects for secondary school completion regression. We find that the likelihood of the beneficiary female to complete secondary school increases by 6.8 percentage points because of the intervention. If we consider the sample of all females in the age-group of 16 to 27 years in Haryana and Punjab irrespective of survey year and re-estimate secondary school completion probability the effect remains positive and significant at 1% level of significance. For estimating the effect on higher secondary completion we consider the sample of women in the age group of 18 to 25 years interviewed in 2015 and 19 to 26 years interviewed in 2016. Nevertheless, the effect on completion of higher secondary education is found to be insignificant (column 3). Considering a sample of all females in the age-group of 18 to 25 years irrespective of survey year also yields insignificant effect (column 4). The absence of any discernible effect on higher secondary education provides evidence that ABAD has been effective in raising the education levels of the beneficiaries but only in a limited capacity up to secondary education.[25] Note that the conditionality here was attached to completion of primary and then the $8^{th}$ standard. In that respect, the program has been successful in improving the chances of completion of secondary schooling, which was not tied to any conditionality.

[Table 4 here]

6.2 *Time allocation*

The TUS we used to gauge the labor market impact of the ABAD allows us to explore daily time allocation of the potential beneficiaries. Despite not going outside and working, are they utilizing the marginal gains in years of education because of the program by spending more time in leisure, socializing or self-care? Or is it the case that the gains got subdued in the entrenched patriarchal norms and the gendered expectations of the households and society at large? To answer these questions, we consider six relevant measures of daily time use: share of time allocated to (i) domestic work (Unpaid domestic services for household and family members), (ii) care-giving (Unpaid caregiving services for household and family members), (iii) socializing (Socializing and communication, community participation and religious practice), (iv) leisure (Culture, leisure, mass-media and sports practices), (v) self-care (Self-care and maintenance) and

(vi) unpaid volunteering work (Unpaid volunteer, trainee and other unpaid work). These six indicators constitute our outcome variables.

As indicated earlier, we use DD regressions as given in equation (3) to estimate the impact on time allocation. Before running the above DD regression, we ensured that we are unable to reject the parallel trends assumption through no statistically significant differences among the elder cohort of the females that is systematic in Haryana in comparison to Punjab (equation 4).[26] Figure 4 presents the marginal effects on the time allocation for the activities listed. As one can observe, we fail to find any discernible changes in the time use patterns of a female beneficiary because of the program. This tends to underscore how deeply rooted the norms on gender and their household roles are and even if the levels of education improved marginally, that got largely subdued.

[Figure 4 here]

6.3 *Is there a delay in marriage?*

To delve further on this, we now consider different aspects of marital outcomes that are often linked to labor participation. For instance, there are multiple ways through which early marriage can act as a deterrent to participation in the labor market (Dhamija and Roychoudhury, 2020). First, it becomes a stumbling block to the pursuance of formal education and acquisition of labor market skills (Field and Ambrus, 2008). Secondly, early marriage raises the opportunity cost to a woman's time spent at home and allocated to domestic chores and her traditional role as caregiver. This becomes even more important with early motherhood which often is linked with early marriage. Hence for the younger brides, the utility of home production increases which reduces her likelihood of work participation (Wang and Wang, 2017). Finally, early marriage is often considered to be linked with the transmission of traditional beliefs that discourage female from going out for work especially in societies plagued with regressive and entrenched patriarchal norms (Asadullah and Wahhaj, 2019). Therefore, it becomes important to gauge the marriage market implication of the ABAD program which potentially can explain the insignificant impact on labor market participation for the beneficiaries despite attaining higher educational levels.

One of the main objectives of the ABAD program was to reduce child marriages and to avail cash benefits required beneficiary females should be unmarried when they turn 18 years. In section 5, we discuss the possibility of a reduction in the prevalence of child marriage because of the program. We formally test this using the sample of 6,677 married females in the eligible and elder age group from Punjab and Haryana belonging to the targeted social group or BPL families from NFHS-4 dataset. The survey collects information on the age at marriage for the sampled household members, which we exploit in this section to study the marital implications. We consider *child marriage* as a binary variable that takes the value of one if the age at marriage is less than 18 years and zero otherwise. To gauge the causal effect of the program, we estimate a DD probit model using all females given by equation (3) with the $Y_{ihds}$ being a dummy variable that takes the value of 1 if the female got married before the age of 18 years and 0 otherwise.

The regression results, presented in table 5 indicate that the treated females in Haryana are found to be 5 percentage points less likely to get married before attaining 18 years of age owing to the intervention.[27] This is in congruence with the large fall in child marriages in Haryana for the treated group that we pointed out in section 5. Next, we estimate the program effects on the likelihood of marriage after turning 18 years using similar regression framework for sample of women in the age-group of 18-25 years and 19-26 years if the year of interview was 2015 and 2016 respectively. This analysis is based on a sample of 3,634 married females for whom the age at marriage is years is at least 18 years. The marginal effects from the DD regression are found to be positive but statistically insignificant (column 2).[28] We also re-estimate the regression for all non-child brides in Haryana and Punjab in the age group of 18 to 25 years irrespective of the survey year. For this sample as well, the double interaction term is insignificant even at 10% level of significance (Column 3). Columns 4 reports the program effects on the probability of marrying at the age of 19 using a sample of 19-24 year old females surveyed in 2015 and 20 to 25 year old females surveyed in the year 2016, who are married but did not marry till 18 years. The findings indicate a 12 percentage higher probability of beneficiary females getting married once they turn 19 compared to control group. In column 5 we report the probability of marrying at 19 using a sample of all females in the age group of 19 to 24 years irrespective of survey year. The double interaction term is still positive but statistically significant at 10% level only. Collectively, our results indicate no gains in age at marriage after turning 18 years; in fact the probability of marriage for the beneficiaries appears to be higher when they turn 19 years. .

[Table 5 here]

Importantly, parents receive the main cash transfer component after their daughters turn 18 years and hence we observe no further incentive for them to move away from the marital norms and delay the marriage of daughters. Our results indicate no apparent evidence that the ABAD program has altered these norms and delayed the age at marriage of targeted females. These findings are in consonance with Das and Nanda (2016), who document about the failure of the cash incentives from the program to improve the age at marriage further beyond 18 years. In fact, they found that the cash was often utilized for meeting marriage expenses. Accordingly, it is possible that the parent's incentive for investing in education of daughters, which is manifested by the significant increase in number of years of education, could be to improve their marriage market prospects of their daughters. It is possible that parents might put higher weightage on non-market returns of education especially in regions with regressive gender norms surrounding marriage. For instance, educating daughters helps in finding suitable grooms and arrange their marriage into a "good" family (Desai and Andrist, 2010; Chiappori *et al.* 2015, Attanasio and Kaufmann, 2017; Adams and Andrew, 2019) Our findings imply the program may not have improved these norms which seem highly entrenched and sticky and the inability to raise the marriage age beyond 18 years can be a possible reason why no effect is observed on female labor market participation.

Importantly the findings also signify that the incentive offered by the program for marrying later at 20 years or 22 years did not have much bearing on the marital age.[29] Therefore this indicates that the marginal disutility of marrying later after 18 years and going against prevailing social norms associated with marriage timing may increase non-linearly and cannot probably be offset by the amount offered. Here it is worth noting that many of the beneficiaries were reported not to have obtained full monetary benefits when they tried redeeming the promised amount (Das and Nanda, 2016). It is possible that these incidents could have led to a trust deficit against the authorities which can potentially be one of the reasons why limited effect on age at marriage are observed.

*Age at first birth*

Despite no significant gains in the age at marriage because of the program, we additionally examine the program implications on fertility. Accordingly, we consider the age at first birth for the sample of married women who married after attaining the legal age of 18 years as our dependent variable and run a DD regression as elucidated in equation 3. We observe no difference in the outcome variable for beneficiaries suggesting that the program may not have improved the onset of fertility among beneficiaries. This is also found to be insensitive to whether we use the survey year to identify the beneficiary and control group or the sample of 18-25 years old married females from Haryana and Punjab respectively. (columns 6 and 7). This again appears to indicate no significant role of the program in altering the rigid and traditional gender norms post marriage despite improvements in education.

*Empowerment*

In relation to gendered norms, the nature of the questions posed in the NFHS-4 questionnaire also allows us to additionally understand the gender relations post marriage for the ABAD beneficiaries relative to others. With higher gains in education because of the program, it is possible that there is a relative improvement in female autonomy within household even when the gains do not get translated into better labor market prospects. The survey administers information on questions that measure female autonomy based on her freedom of mobility, say in the household and husband's attitude towards wife among others. Appendix A8 lists the questions and the variable definitions that include six questions largely capturing measures on empowerment. To estimate the role of the program on changing these empowerment or attitude measures, we again use the DD regression framework as chalked out in equation 3. Figure 5 presents the marginal effects from the DD regression obtained with these six empowerment measures as outcome variables. The findings largely indicate no improvement in these empowerment measures because of the intervention further lending credence to our findings of no substantial changes in norms on account of the ABAD program. These findings resonate with that by Litwin et al. (2019) that document the ineffectiveness of the Bolsa Familia program in reducing extreme intimate partner violence events in Brazil.

[Figure 5 here]

## 7. Discussion and Conclusion

Normal educational CCTs make periodic payments to beneficiaries in lieu of satisfying certain conditions that include getting enrolled in school or ensuring school attendance among others. On the contrary, certain others programs depend on a payment "promise" or long term financial incentives for satisfying limited number of conditions with an expectation that the benefits can be derived in the process. In this paper, we evaluate the indirect effects of one such program called the ABAD implemented in the Indian state of Haryana. The program promises monetary incentives redeemable only after the daughters turn 18 years and remain unmarried to poor families and those belonging from socially deprived groups. In particular, we gauge the educational and labor supply decision of the women beneficiaries, and in the process of identifying the channels for our empirical findings, we also assess the program effects on time-use, age at marriage, age at first birth and post-marital indicators of empowerment and bargaining power. We call these as "indirect outcomes" because the conditionality of the program is not linked to any of these but only remaining unmarried till 18 years of age.

We evaluate these program effects using multiple representative datasets wherein the exogenous timing of the program that spanned from 1994 to 1998 is exploited. The findings indicate significant reduction in child marriage as a direct effect of the conditionality attached. In terms of the indirect effects, we find significant gains in years of education that led to higher completion rate of secondary education among beneficiaries though no discernible effect on labor participation is observed. These effects are largely equal among beneficiaries of different social groups or sectors. While gauging the potential channels or reasons for these observations, we find the no impact of the program in raising educational attainment beyond secondary education. Further, no difference in marriage at 18 treated and control group females is observed though the likelihood of marrying at 19 years is found to increase marginally because of the program with no impact on age at first child birth. Additionally, the beneficiaries are not more likely to spend time on leisure, socializing or self-care activities and no indication of improvement in indicators of empowerment and bargaining power is observed. This set of evidence indicates that the program failed in altering the prevailing gender norms despite improvements in educational outcomes.

The paper is pertinent since it is among the first work to provide evidence on protracted benefits to the parents. We provide evidence of the success of such monetary "promise" of long term payment to incentivize parents not to marry off their daughters early and utilize the time before marriage by investing in human capital. In this context, existing programs like Laadli Laxmi Scheme or the Girl Child Protection Scheme implemented in parts of India and even in other developing countries that include the Female Secondary School Assistance Project in Bangladesh and the Female School Stipend Program in Pakistan among others have significant importance for the stakeholders. Empirical evidence on evaluation of these programmes documents significant reduction of child marriage along with considerable gains on educational outcomes (Malhotra and Elnakib, 2021).

Nevertheless, the paper also documents evidence on outcomes where the program failed to produce any discernible effect. For instance, despite the reduction in child marriage and an increase in the years of education leading to higher completion of secondary education among beneficiaries, the gains did not translate into higher labor participation. In fact, age at marriage after 18 years of age, time allocation for leisure of self-care or even age at first birth did not show an increase. Further, no improvements in the indicators of empowerment or fertility age point out to the fact that the entrenched and sticky gender norms did not get altered because of the intervention. Prevailing gender norms that mandate women with traditional roles as caregivers inhibit the initial gains through the program. Accordingly, complementary interventions that can change the prevailing norms need to be complemented with the financial incentives to improve the status of females in the household and society. In this respect, direct actions related to awareness generation or projection of role models that can improve the normative gender beliefs can be helpful. Additionally, *norms messaging* that can shape up or alter social expectations surrounding gender roles can prove to be a necessary precursor to social change (Bicchieri, 2018). Empirical evidence in the context of exclusive breastfeeding in Mali, sanitation practice in India and female labor participation in Saudi Arabia among others have documented potential benefits through targeting social expectations by dissemination of behavioral messages that bring to the forefront the emergence of a new norm and the decline of the old regressive norm (for example, non-exclusive breastfeeding feeding or open defecation or not allowing women to work) (Bicchieri et al. 2021; Ashraf et al. 2020; Burztyn et al. 2020). Importantly, the shift in social expectations and beliefs about other's behavior may become self-reinforcing as more people will then potentially adopt the target behavior and make the behavior more sustainable ensuring reduction in women's withdrawal from labor market due to traditional gender norms.

Nevertheless, the effectiveness of these interventions needs to be tested and this can be an agenda for further research where experimental evidence can be generated.

Further, because we did not observe any gains in labor supply decisions for women beneficiaries despite an increase in education, incentives to encourage their participation can additionally be considered that can complement interventions like ABAD. Here policies revolving around incentivizing female labor participation from the household in similar veins to educational or health CCT can be considered. Similar to cash incentives given to parents for sending their daughters to schools or getting regular health check-ups, interventions that give direct or indirect transfers to poorer families for sending women to work can be thought of as a policy instrument to encourage their work participation. Additionally, interventions to encourage women to acquire higher education and skill development required for effective participation in the labor market need to be strengthened. Nevertheless, as was in the earlier case, this requires to be tested to gauge its effectiveness and hence can constitute future research agenda.

# **Figures**

Figure 1: Educational attainment of females in Haryana

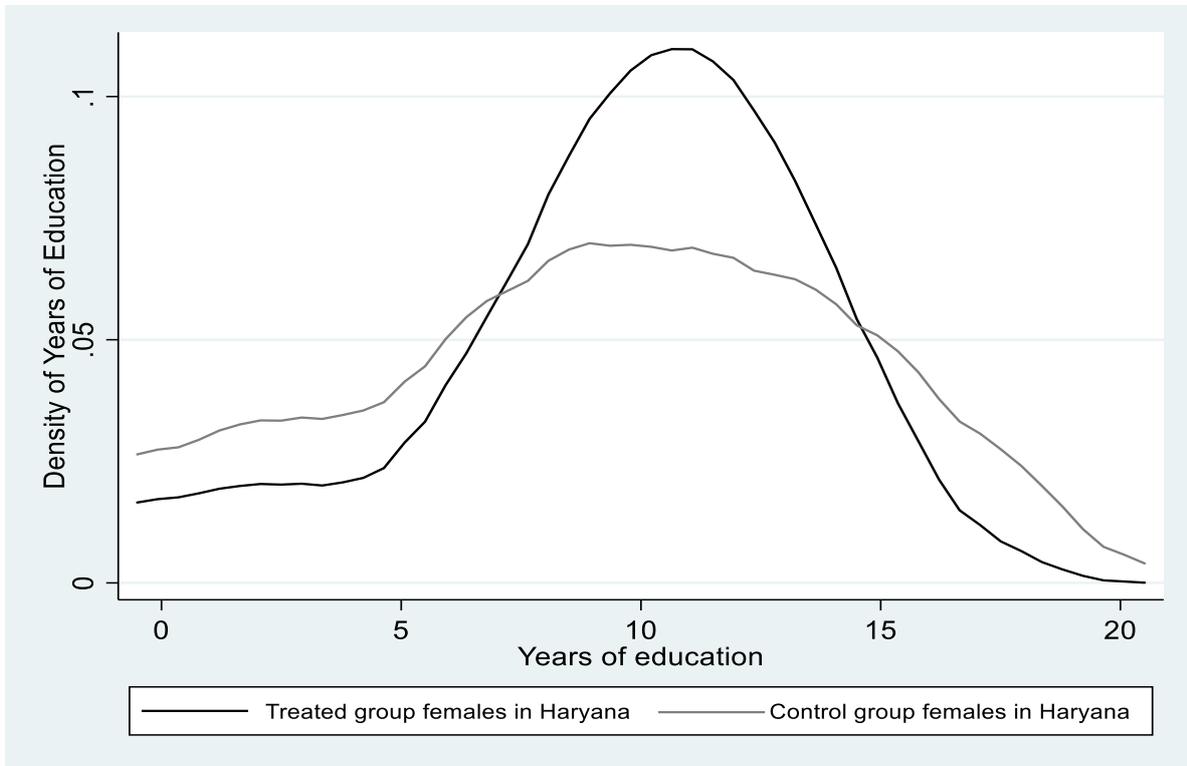

Note: Figure presents the kernel density of years of education of females in treated and control age cohort in Haryana.

Figure 2: Heterogeneous effects of ABAD

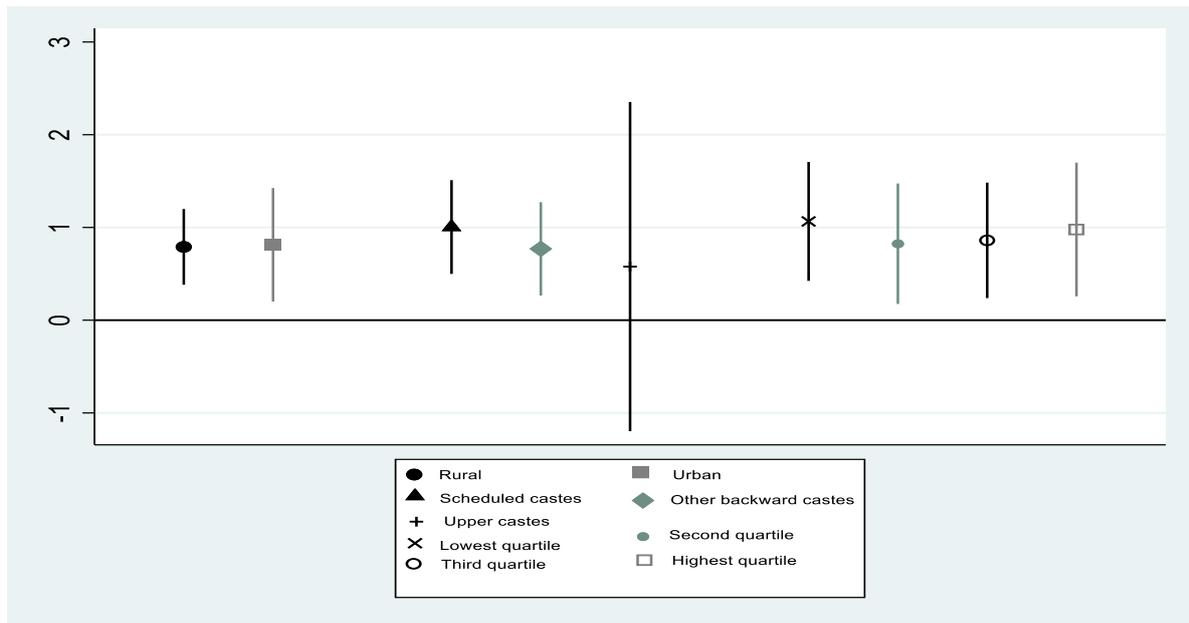

2(a) : Effect of ABAD across sector, caste groups and wealth quartiles

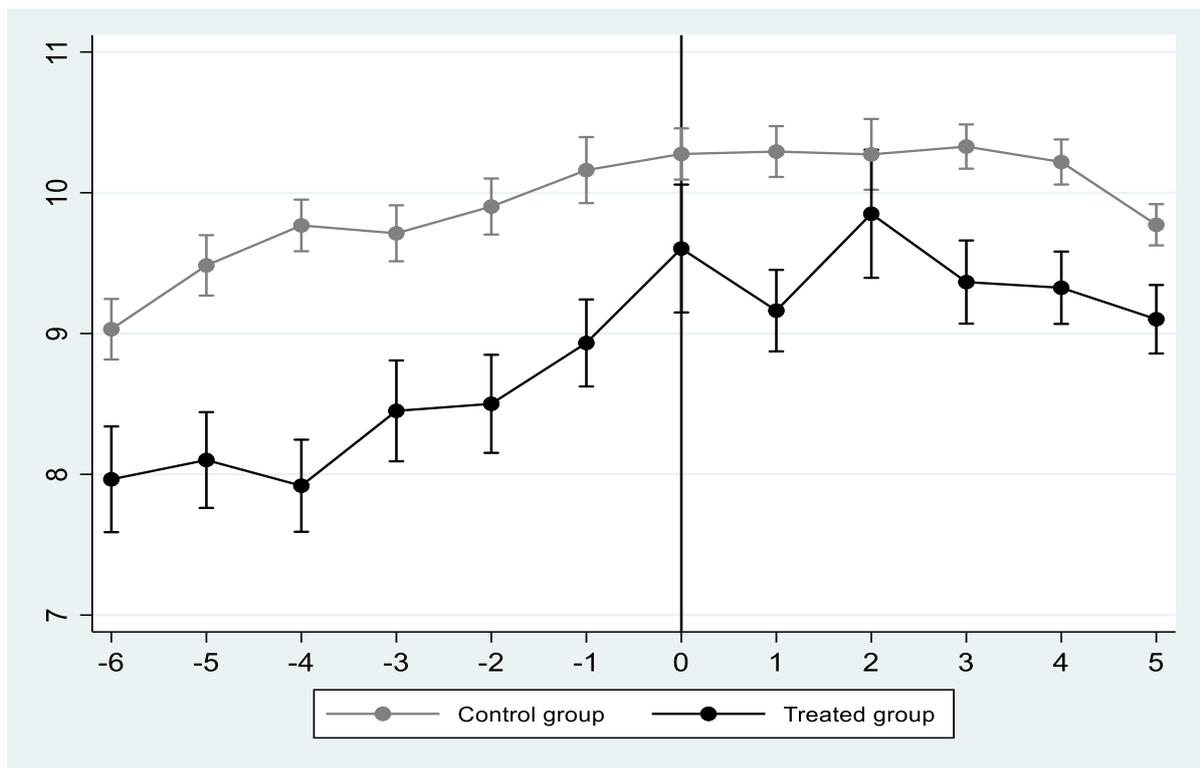

2(b) : Effect of ABAD over time

Note. The average marginal effects from OLS regression are presented along with 95% confidence interval. All regressions control for the covariates including household size, religion, wealth score, current age and area of residence. The standard errors are calculated by clustering at the primary sampling unit level.

Figure 4: Effect of ABAD on share of time allocation

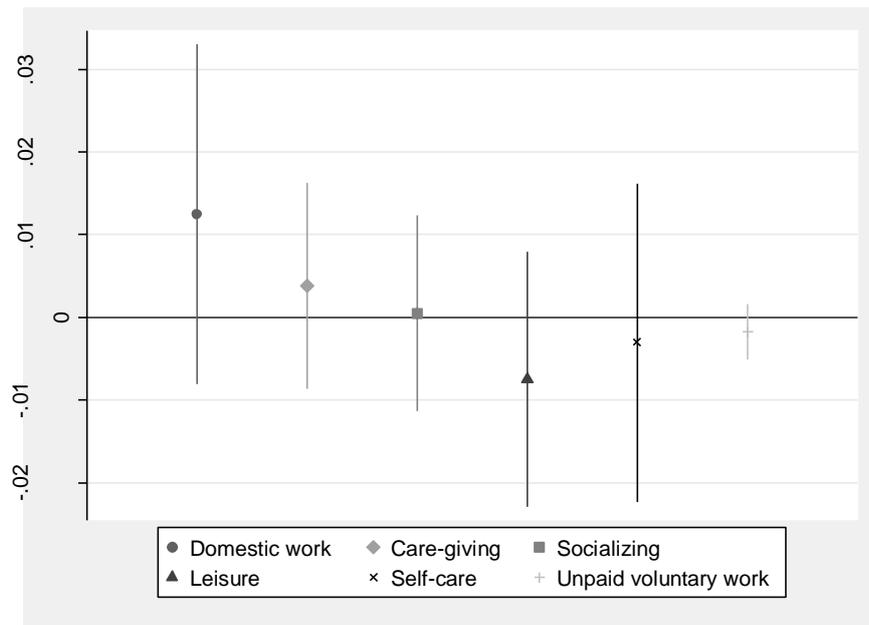

Note. The marginal effects from probit regression are presented along with 95% confidence interval. All regressions control for the covariates including household size, religion, wealth score, current age and area of residence. The standard errors are calculated by clustering at the primary sampling unit level.

Figure 5: Effect of ABAD on empowerment of married women

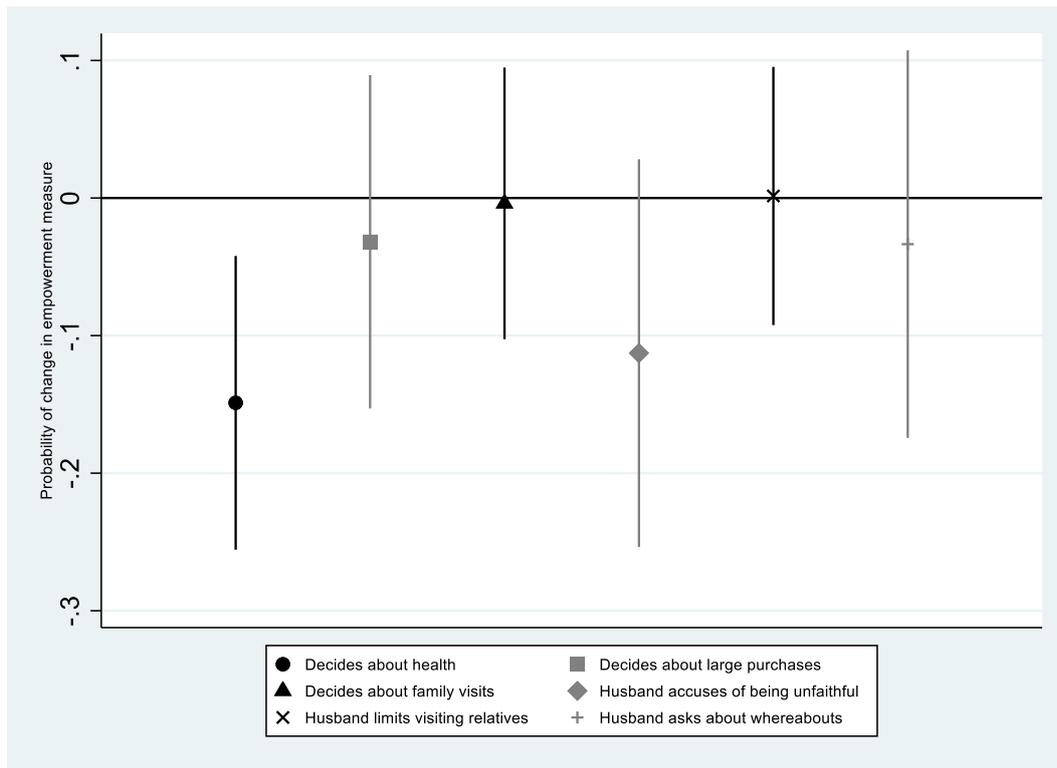

Note. The marginal effects from probit regression are presented along with 95% confidence interval. All regressions control for the covariates including household size, religion, wealth score, current age and area of residence. The standard errors are calculated by clustering at the primary sampling unit level.

**Tables**

Table 1: Descriptive statistics of females in Haryana in the age-cohort of 16-28years

|  | All | Treated group | Control group |
|---|---|---|---|
| *Panel A: NFHS-4 (2015-16)* | | | |
| Education (in years) | 9.76 | 9.78 | 9.73 |
| Age (in years) | 21.56 | 18.53 | 24.37 |
| HH size | 5.93 | 5.87 | 6.00 |
| Hindus (share in %) | 87.68 | 85.24 | 88.77 |
| Wealth score | 78429.64 | 70285.23 | 84575.71 |
| Rural (share in %) | 66.18 | 67.32 | 65.12 |
| N | 16,439 | 8,001 | 8,438 |
| *Panel B: PLFS (2017-18) and PLFS (2019)* | | | |
|  | All | Treated group | Control group |
| Working (share) | 0.11 | 0.07 | 0.15 |
| Skilled labor (share) | 0.09 | 0.06 | 0.12 |
| Age (in years) | 24.36 | 21.56 | 27.43 |
| HH size | 5.60 | 5.52 | 5.68 |
| Hindus (share in %) | 88.82 | 88.02 | 89.69 |
| Household MPCE (in Indian currency rupees) | 2518.05 | 2615.33 | 2411.49 |
| Rural (share in %) | 52.85 | 51.63 | 54.17 |
| N | 2,460 | 1,286 | 1,174 |
| *Panel C: TUS (2019)* | | | |
|  | All | Treated group | Control group |
| Working outside (share in time devoted every day) | 0.03 | 0.03 | 0.04 |
| Age (in years) | 25.52 | 22.75 | 28.20 |
| HH size | 5.00 | 5.00 | 4.99 |
| Hindus (share in %) | 89.97 | 89.53 | 90.38 |
| Household MPCE (in Indian currency rupees) | 3015.49 | 2904.28 | 3122.78 |
| Rural (share in %) | 59.54 | 64.12 | 55.13 |
| N | 1226 | 602 | 624 |

*Note: For PLFS and TUS, the statistics for all females from households from the bottom 25 percentile of Haryana or belonging to the Scheduled Caste (SC) or Scheduled Tribe (ST) groups are given.*

Table 2: Effect of ABAD on education

| | Treatment effects | | | Parallel trends | |
|---|---|---|---|---|---|
| | Without controls | With controls | Age cohort- 22-28year | Age cohort- 22-29year | Age cohort- 22-32year |
| | (1) | (2) | (3) | (4) | (5) |
| Female*Age-eligible*Haryana | 0.734*** | 0.808*** | | | |
| | (0.204) | (0.174) | | | |
| Female*Haryana*Age | | | 0.130 | 0.057 | 0.053 |
| | | | (0.079) | (0.064) | (0.040) |
| Female* Age-eligible | 0.245 | 0.408*** | | | |
| | (0.151) | (0.129) | | | |
| Female | -0.143 | -0.269** | 3.154** | 2.803** | 2.382*** |
| | (0.123) | (0.106) | (1.470) | (1.211) | (0.748) |
| Age group | 0.558*** | 0.437*** | | | |
| | (0.108) | (0.119) | | | |
| Haryana | 1.632*** | 1.172*** | 2.939* | 3.459*** | 2.351*** |
| | (0.153) | (0.313) | (1.553) | (1.289) | (0.892) |
| Female*Haryana | -1.568*** | -1.504*** | -4.819** | -3.077* | -2.945*** |
| | (0.165) | (0.143) | (1.993) | (1.633) | (1.061) |
| Age-eligible *Haryana | -1.025*** | -0.970*** | | | |
| | (0.143) | (0.126) | | | |
| Age | | -0.064*** | -0.164*** | -0.094*** | -0.142*** |
| | | (0.013) | (0.031) | (0.032) | (0.020) |
| Female*Age | | | -0.135** | -0.122*** | -0.106*** |
| | | | (0.058) | (0.046) | (0.027) |
| Haryana*Age | | | -0.064 | -0.087* | -0.042 |
| | | | (0.056) | (0.045) | (0.029) |
| Controls | N | Y | Y | Y | Y |
| District FE | N | Y | Y | Y | Y |
| Observations | 27,390 | 27,390 | 13,953 | 16,080 | 21,029 |
| R-squared | 0.020 | 0.295 | 0.357 | 0.360 | 0.371 |

The coefficients are the average marginal effect. All regressions control for the covariates including household size, religion, wealth score, current age, area of residence and district fixed effects. Standard errors are depicted in the parentheses and clustered at the primary sampling unit level. FE indicates fixed effects.
*** $p<0.01$, ** $p<0.05$, * $p<0.1$

Table 3: Effects on work participation

| | Treatment effects | | | Parallel trends | | |
|---|---|---|---|---|---|---|
| | PLFS | | TUS | PLFS | | TUS |
| | Working | Skilled work | Working | Skilled work | Working | Skilled work |
| Haryana*Female*Age eligible | -0.115* | -0.082 | 0.002 | | | |
| | (0.068) | (0.082) | (0.018) | | | |
| Haryana*Female*Age | | | | -0.010 | -0.025 | -0.005 |
| | | | | (0.020) | (0.026) | (0.007) |
| Haryana*Female | 0.026 | -0.096* | 0.021* | 0.593 | -0.149 | 0.178 |
| | (0.050) | (0.050) | (0.012) | (0.703) | (0.533) | (0.210) |
| Haryana*Age eligible | 0.020 | -0.042 | 0.000 | | | |
| | (0.044) | (0.038) | (0.015) | | | |
| Female*Age eligible | 0.225*** | 0.050 | 0.069*** | | | |
| | (0.041) | (0.043) | (0.013) | | | |
| Haryana*Age | | | | 0.008 | 0.015 | 0.010* |
| | | | | (0.015) | (0.015) | (0.006) |
| Female*Age | | | | -0.016 | 0.008 | 0.000 |
| | | | | (0.014) | (0.017) | (0.005) |
| Haryana | -0.019 | 0.151** | -0.076*** | -0.268 | -0.248 | -0.391** |
| | (0.072) | (0.076) | (0.026) | (0.427) | (0.449) | (0.182) |
| Female | -0.565*** | -0.286*** | -0.238*** | -0.049 | -0.536 | -0.242 |
| | (0.024) | (0.026) | (0.008) | (0.391) | (0.463) | (0.153) |
| Age eligible | -0.065* | 0.050 | -0.026** | | | |
| | (0.037) | (0.036) | (0.012) | | | |
| Age | 0.032*** | 0.025*** | 0.008*** | 0.036*** | 0.018* | 0.002 |
| | (0.004) | (0.004) | (0.001) | (0.011) | (0.009) | (0.004) |
| Control | Yes | Yes | | Yes | Yes | |
| District FE | Yes | Yes | | Yes | Yes | |
| Survey quarter and round FE | Yes | Yes | | Yes | Yes | |
| Observations | 2,070 | 2,070 | 3,156 | 1,025 | 1,017 | 1,510 |

The coefficients are the average marginal effect. All regressions control for the covariates including household size, wealth score, current age, area of residence, average per-capita monthly expenditure, survey and district fixed effects. Standard errors are depicted in the parentheses and clustered at the primary sampling unit level. FE indicates fixed effects. Females from households in the lowest 25 percentile in respective states or those from the SC or OBC social group are considered in the sample.
*** p<0.01, ** p<0.05, * p<0.1

Table 4: Effect of ABAD on secondary completion and higher education

| | Secondary education | | Higher secondary education | |
|---|---|---|---|---|
| | Considering survey year[+] | Not considering survey year[@] | considering survey year[#] | Not considering survey year[^] |
| Female*Age eligible*Haryana | 0.068*** | 0.074*** | 0.034 | 0.037 |
| | (0.021) | (0.022) | (0.022) | (0.023) |
| Female*Age eligible | 0.058*** | 0.050*** | 0.026 | 0.020 |
| | (0.016) | (0.017) | (0.017) | (0.018) |
| Haryana*Age eligible | -0.155*** | -0.129*** | -0.087*** | -0.051*** |
| | (0.015) | (0.016) | (0.016) | (0.016) |
| Female *Haryana | -0.124*** | -0.131*** | -0.098*** | -0.102*** |
| | (0.015) | (0.016) | (0.015) | (0.016) |
| Age eligible | 0.175*** | 0.180*** | 0.074*** | 0.081*** |
| | (0.015) | (0.015) | (0.016) | (0.016) |
| Female | -0.017 | -0.009 | 0.026** | 0.033** |
| | (0.012) | (0.013) | (0.011) | (0.013) |
| Haryana | 0.146*** | 0.113*** | 0.163*** | 0.156*** |
| | (0.031) | (0.033) | (0.033) | (0.033) |
| Age | 0.025*** | 0.030*** | 0.014*** | 0.024*** |
| | (0.002) | (0.002) | (0.002) | (0.002) |
| Controls | Y | Y | Y | Y |
| District FE | Y | Y | Y | Y |
| Observations | 27,386 | 26,446 | 19,030 | 19,038 |

The coefficients are the average marginal effect. All regressions control for the covariates including household size, social group, wealth score, current age, area of residence and district fixed effects. Standard errors are depicted in the parentheses and clustered at the primary sampling unit level. FE indicates fixed effects.
*** p<0.01, ** p<0.05, * p<0.1
[+]Sample of treated and control individuals in the age cohort of 16-27 years interviewed in 2015 and in the age cohort of 17-28 years interviewed in 2016.
[@] Sample of treated and control individuals in the age cohort of 16-27 irrespective of survey year.
[#] Sample of treated and control individuals in the age cohort of 18-25 years interviewed in 2015 and in the age cohort of 19-26 years interviewed in 2016.
[^]Sample of treated and treated and control individuals in the age cohort of 18-25 irrespective of survey year.

Table 5: Effect of ABAD on marriage practices and fertility

| | (1) Child marriage | (2) Marrying at 18[+] | (3) Marrying at 18[^] | (4) Marrying at 19[@] | (5) Marrying 19[#] | (6) Age at first birth[+] | (7) Age at first birth[^] |
|---|---|---|---|---|---|---|---|
| Haryana*Age eligible | -0.050** | 0.020 | -0.012 | 0.122*** | 0.072* | 0.034 | 0.015 |
| | (0.025) | (0.028) | (0.030) | (0.041) | (0.044) | (0.115) | (0.116) |
| Haryana | 0.136*** | 0.013 | 0.024 | 0.027 | -0.036 | -0.566* | -0.512* |
| | (0.048) | (0.064) | (0.075) | (0.088) | (0.086) | (0.295) | (0.299) |
| Age eligible | 0.101*** | -0.000 | -0.005 | -0.110** | -0.124*** | -0.196 | -0.201 |
| | (0.025) | (0.032) | (0.033) | (0.046) | (0.047) | (0.125) | (0.123) |
| Age | -0.000 | -0.042*** | -0.054*** | -0.083*** | -0.110*** | 0.404*** | 0.398*** |
| | (0.003) | (0.006) | (0.007) | (0.011) | (0.012) | (0.028) | (0.029) |
| Controls | Y | Y | Y | Y | Y | Y | Y |
| District FE | Y | Y | Y | Y | Y | Y | Y |
| Observations | 6,677 | 3,634 | 3,356 | 2,176 | 1,951 | 2,428 | 2,173 |
| R-squared | | | | | | 0.246 | 0.233 |

The coefficients are the average marginal effect. All regressions control for the covariates including household size, religion, wealth score, current age, area of residence and district fixed effects. Standard errors are depicted in the parentheses and clustered at the primary sampling unit level. FE indicates fixed effects.

*** p<0.01, ** p<0.05, * p<0.1.

[+] Sample of treated and control females in the age cohort of 18-26 years based on survey year who were not married before turning 18

[^] Sample of treated and control females in the age cohort of 18-25 years irrespective of survey year who were not married before turning 18

[@] Sample of treated and control females in the age cohort of 19-25 years based on survey year who were not married before turning 19

[#] Sample of treated and control females in the age cohort of 19-25 years based on survey year who were not married before turning 19

# Appendix

## Table A1: Variables and definition

| Variable | Description |
| --- | --- |
| *From NFHS-4 (2015-16)* | |
| *Outcome variables* | |
| Education | Years of education |
| Secondary completion | Dummy variable=1 for individuals who have completed secondary schooling and zero for others |
| Higher education | Dummy variable=1 for individuals who have more than secondary schooling and zero for others |
| Child marriage | Dummy variable=1 for individuals who were married before attaining 18 year and zero for other married women |
| Married at 18 | Dummy variable=1 for individuals who got married on turning 18 year and zero for other married women |
| Married at 19 | Dummy variable=1 for individuals who got married on turning 19 year and zero for other married women |
| Age at first birth | Age of the married woman (in years) when first child was born |
| *Interest variables* | |
| Female | Dummy=1 for females and zero for males |
| Haryana | Dummy=1 for individuals residing in Haryana and zero for those residing in Punjab |
| Age eligible | Dummy variable=1 if the individual is in the age group of 18-22years and zero otherwise |
| Female*Haryana | 1 for females from Haryana and zero for others |
| Age eligible*Haryana | 1 for all individuals in the age group of 18-22 years from residing in Haryana and zero for others |
| Female*Age eligible | 1 for females in the age group of 18-22 years and zero for others |
| Female*Age eligible *Haryana | 1 for females in the age group of 18-22 years residing in Haryana and zero for others |
| *Other controls* | |
| HH size | Number of household members |
| Wealth score | Wealth score based on assets held by the households |
| Religion | Categorical variable identifying the religious affiliation of the individual as Hindu, Muslim, |
|  | Christian or others |
| Caste category | Categorical variable identifying the caste group of the individual as SC, ST, OBC or general |
| Rural | Dummy variable=1 if the individual is from rural areas and zero for those residing in urban areas |

*From PLFS (2017-18) and (2018-19)*

| | |
|---|---|
| *Outcome variables* | |
| Labor supply (working) | Dummy variable=1 for individuals who reported working in household enterprize, as unpaid family worker, as regular salaried/ wage employee, as casual wage labour: in public works or in other types of works as "Usual Principal Status" (The activity status on which a person spent relatively long time (major time criterion) during the 365 days preceding the date of survey) and zero otherwise |
| Labor supply (skilled working) | Dummy variable=1 for individuals who reported working in household enterprize, as unpaid family worker, as regular salaried/ wage employee as "Usual Principal Status" and zero otherwise |
| *Interest variables* | |
| Female | Dummy=1 for females and zero for males |
| Haryana | Dummy=1 for individuals residing in Haryana and zero for those residing in Punjab |
| Age group | Dummy variable=1 if the individual is in the age group of 18-23 years for those surveyed in 2017; 19-24 years for those surveyed in 2018 and 20-25 for those surveyed in 2019 and zero for those in the age cohort 24-29 years surveyed in 2017, 25-30 years in 2018 and 26-31 years in 2019 |
| Female*Haryana | 1 for females from Haryana and zero for others |
| Age eligible*Haryana | 1 for all individuals in the age group of 1 from residing in Haryana (indicated above) and zero for others |
| Female*Age eligible | 1 for females in the age group of 1 (indicated above) and zero for others |
| Female*Age eligible *Haryana | 1 for females in the age group of 1 from residing in Haryana (indicated above) and zero for others |
| *Other controls* | |
| HH size | Number of household members |
| MPCE | Monthly per-capita consumption expenditure (in Indian rupees) |
| Religion | Categorical variable identifying the religious affiliation of the individual as Hindu, Muslim, and others |
| Caste category | Categorical variable identifying the caste group of the individual as SC, ST, OBC or general |
| Rural | Dummy variable=1 if the individual is from rural areas and zero for those residing in urban areas |
| Round | Dummy variable=1 for individuals were surveyed round two of PLFS (2018-19) an zero otherwise |
| Quarter | Categorical variable identifying the quarter of the year when the survey was conducted- quarter1, quarter 2, quarter 3 and quarter 4. |

*From TUS (2019)*

*Outcome variables*

| | |
|---|---|
| Labor supply | Share of time allocated to "employment and related activities" and "production of goods for final use" on the day prior to the survey |
| Share of time for domestic activities | Share of time allocated to "unpaid domestic services for household and family members" on the day prior to the survey |
| Share of time for care-giving | Share of time allocated to "unpaid caregiving services for household and family members" on the day prior to the survey |
| Share of time for socializing | Share of time allocated to "Socializing and communication, community participation and religious practice" on the day prior to the survey |
| Share of time for leisure | Share of time allocated to "Culture, leisure, mass-media and sports practices" on the day prior to the survey |
| Share of time for self-care | Share of time allocated to "Self-care and maintenance" on the day prior to the survey |
| Share of time for unpaid volunteering work | Share of time allocated to "Unpaid volunteer, trainee and other unpaid work" on the day prior to the survey |

*Interest variables*

| | |
|---|---|
| Female | Dummy=1 for females and zero for males |
| Haryana | Dummy=1 for individuals residing in Haryana and zero for those residing in Punjab |
| Age group | Dummy variable=1 if the individual is in the age group of 20-25 years and zero if the age is in between 26 to 31 years. |
| Female*Haryana | 1 for females from Haryana and zero for others |
| Age eligible*Haryana | 1 for all individuals in the age group of 20-25 years from residing in Haryana and zero for others |
| Female*Age eligible | 1 for females in the age group of 20-25 years and zero for others |
| Female*Age eligible *Haryana | 1 for females in the age group of 20-25 years residing in Haryana and zero for others |
| *Other controls* | |
| HH size | Number of household members |
| MPCE | Monthly per-capita consumption expenditure (in Indian rupees) |
| Religion | Categorical variable identifying the religious affiliation of the individual as Hindu, Muslim, and others |
| Caste category | Categorical variable identifying the caste group of the individual as SC, ST, OBC or general |
| Rural | Dummy variable=1 if the individual is from rural areas and zero for those residing in urban areas |

Table A2: Effect of ABAD on education- Placebo effects

| | Pseudo treatment | | | |
|---|---|---|---|---|
| | Pseudo treatment-22-28age group | Pseudo treatment-24-30age group | NFHS-3 | Non-BPL and UC |
| Female*Age eligible *Haryana | -0.193 | -0.173 | 0.088 | 0.264 |
| | (0.197) | (0.200) | (0.559) | (0.226) |
| Female | -0.933*** | -1.145*** | -1.450*** | 0.321*** |
| | (0.110) | (0.111) | (0.313) | (0.105) |
| Age eligible | -0.339** | -0.323** | -0.426 | -0.047 |
| | (0.134) | (0.139) | (0.335) | (0.135) |
| Female*Age eligible | 0.601*** | 0.546*** | 0.978** | 0.428*** |
| | (0.144) | (0.153) | (0.387) | (0.132) |
| Haryana | 1.025*** | 0.944*** | 1.262*** | 1.651*** |
| | (0.350) | (0.340) | (0.337) | (0.460) |
| Female*Haryana | -1.295*** | -1.248*** | -1.219*** | -0.998*** |
| | (0.149) | (0.146) | (0.447) | (0.170) |
| Age eligible*Haryana | 0.275* | 0.367*** | -0.093 | -0.282* |
| | (0.141) | (0.141) | (0.372) | (0.164) |
| Age | -0.177*** | -0.175*** | -0.140*** | 0.108*** |
| | (0.015) | (0.015) | (0.033) | (0.019) |
| Controls | Y | Y | Y | Y |
| District FE | Y | Y | N | Y |
| Observations | 24,618 | 23,532 | 3,439 | 11,506 |
| R-squared | 0.376 | 0.405 | 0.363 | 0.241 |

The coefficients are the average marginal effect. All regressions control for the covariates including household size, religion, wealth score, current age, area of residence and district fixed effects. Standard errors are depicted in the parentheses and clustered at the primary sampling unit level. FE indicates fixed effects.
*** $p<0.01$, ** $p<0.05$, * $p<0.1$

Figure A3: Effect of ABAD on years of education using various wealth quintiles using NFHS-4

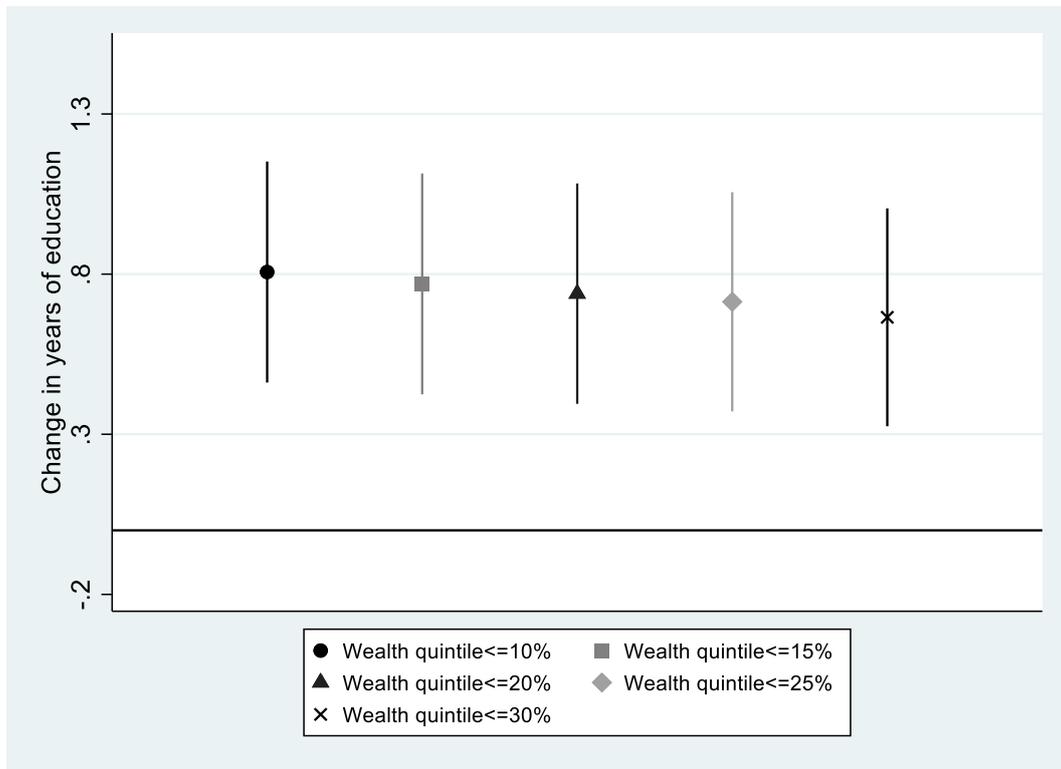

The average marginal effects are plotted along with the 95% confidence intervals. All regressions control for the covariates including household size, wealth score, current age, area of residence, average per-capita monthly expenditure, survey and district fixed effects. Standard errors are clustered at the primary sampling unit level.

Figure A4: Robustness (with different MPCE cut-offs)

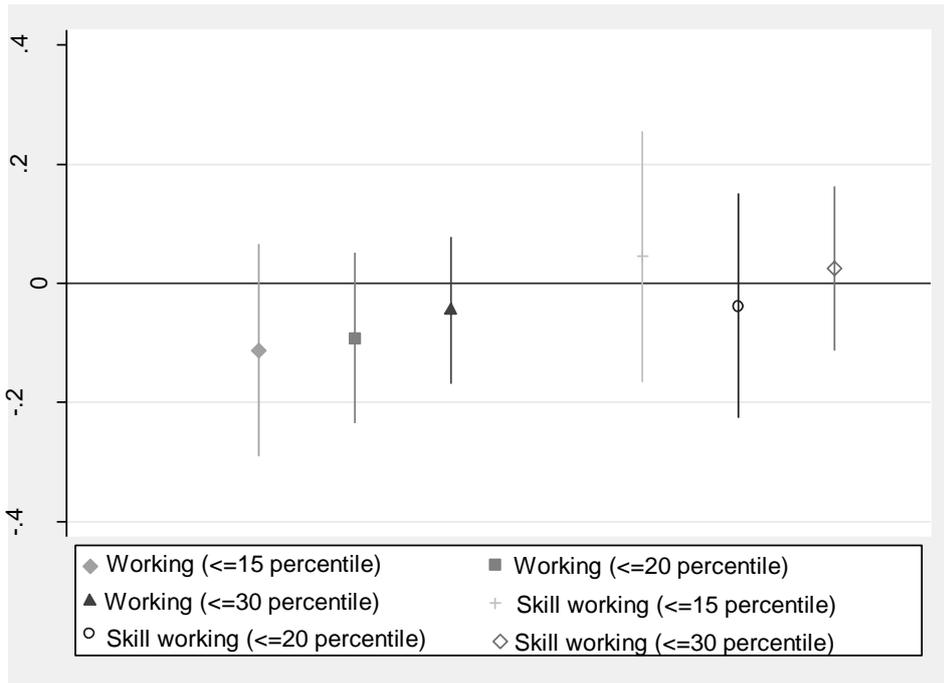

The average marginal effects are plotted along with the 95% confidence intervals. All regressions control for the covariates including household size, wealth score, current age, area of residence, average per-capita monthly expenditure, survey and district fixed effects. Standard errors are clustered at the primary sampling unit level.

Table A5: Effect of ABAD using alternate sample (with neighboring districts)

|  | Punjab and Haryana | Punjab, Himachal Pradesh, Uttarakhand and Haryana | Punjab, Himachal Pradesh, Uttarakhand, Uttar Pradesh and Haryana |
|---|---|---|---|
| Female*Age eligible*Haryana | 0.887*** | 0.612** | 0.719*** |
|  | (0.285) | (0.250) | (0.224) |
| Female | -0.266 | -0.325** | -0.324** |
|  | (0.194) | (0.151) | (0.151) |
| Age eligible | 0.241 | 0.246 | 0.330* |
|  | (0.215) | (0.174) | (0.169) |
| Female*Age eligible | 0.475** | 0.558*** | 0.543*** |
|  | (0.204) | (0.166) | (0.166) |
| Haryana | -0.236 | -0.739** | 0.648* |
|  | (0.343) | (0.359) | (0.354) |
| Female*Haryana | -1.278*** | -1.074*** | -1.677*** |
|  | (0.244) | (0.203) | (0.195) |
| Age eligible*Haryana | -0.840*** | -0.624*** | -0.721*** |
|  | (0.228) | (0.199) | (0.176) |
| Age | -0.087*** | -0.065*** | -0.050*** |
|  | (0.024) | (0.020) | (0.018) |
| Controls | Y | Y | Y |
| District FE | Y | Y | Y |
| Observations | 8,692 | 11,272 | 16,086 |
| R-squared | 0.270 | 0.270 | 0.299 |

The coefficients are the average marginal effect. All regressions control for the covariates including household size, religion, wealth score, current age, area of residence and district fixed effects. Standard errors are depicted in the parentheses and clustered at the primary sampling unit level. FE indicates fixed effects.
*** p<0.01, ** p<0.05, * p<0.1

Figure A6: Parallel trends in Haryana and the neighboring districts

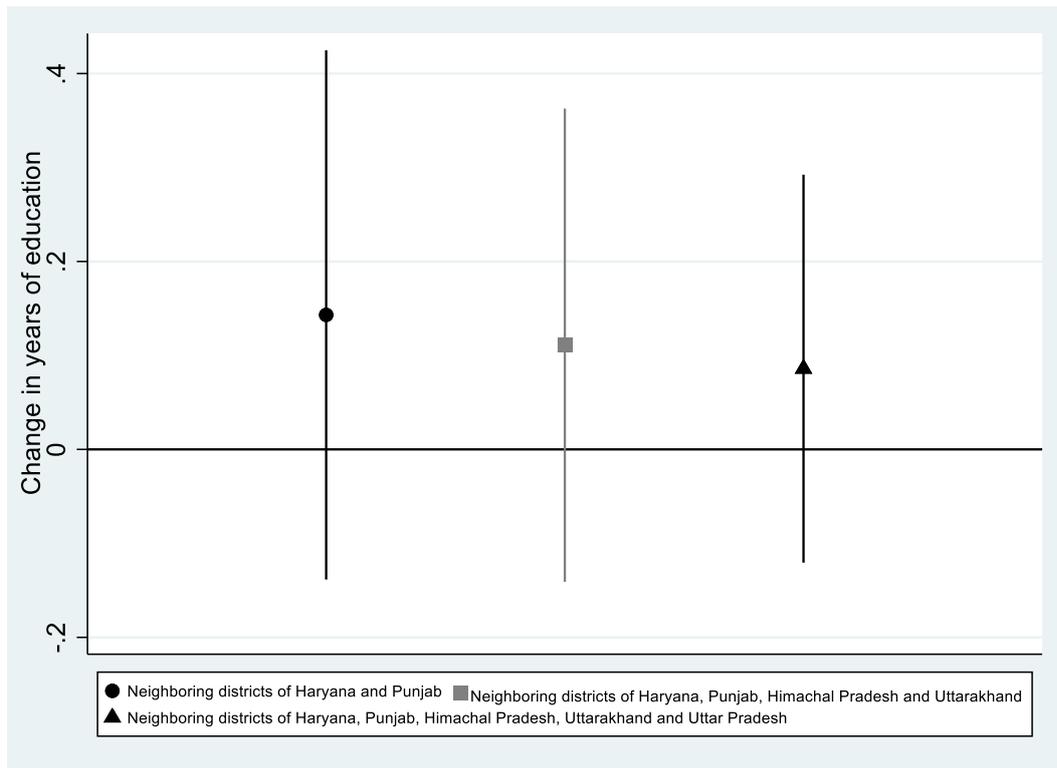

Note. The average marginal effects from OLS regression are presented along with 95% confidence interval. All regressions control for the covariates including household size, religion, wealth score, current age and area of residence. The standard errors are calculated by clustering at the primary sampling unit level.

Figure A7: Alternate definition of age-eligibility

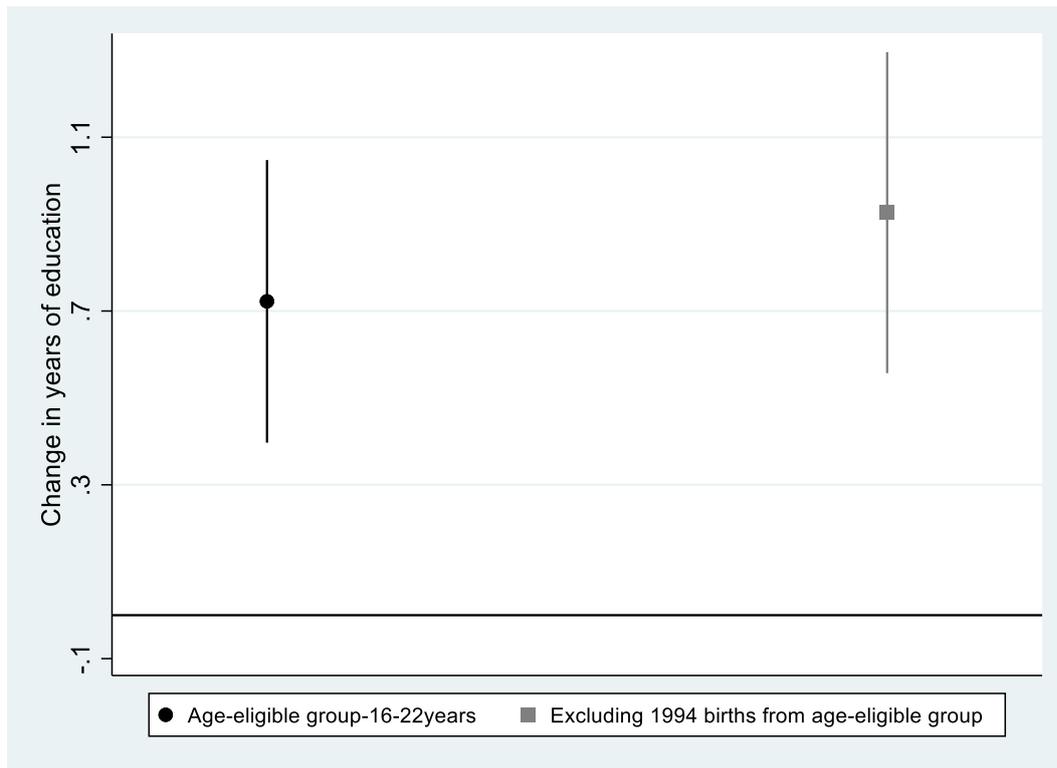

Note. The average marginal effects from OLS regression are presented along with 95% confidence interval. All regressions control for the covariates including household size, religion, wealth score, current age and area of residence. The standard errors are calculated by clustering at the primary sampling unit level.

Table A8: Empowerment measures

| Indicator | Description |
|---|---|
| **Ability to make decisions-** *Each indicator is coded as **1** if the response is "**respondent**" and zero for "**respondent and husband jointly**" or for "**husband**" or "**someone else**"* | |
| | Who usually makes decisions about health care for yourself? |
| | Who usually makes decisions about making major household purchases? |
| | Who usually makes decisions about visits to your family or relatives? |
| **Household relations -** | *Each indicator is coded as **1** if the response is "**no**" and **0** for "**yes**"* |
| | Does he (husband) frequently accuse you of being unfaithful? |
| | Does he (husband) try to limit your contact with your family? |
| | Does he (husband) insist on knowing where you are at all times? |

---

[1] Scheduled Castes and Other Backward Castes are among the historically most deprived social groups in India, both socially as well as economically (Deshpande, 2000; Sundaram and Tendulkar, 2003)

[2] In India, females are said to have entered child marriage if they marry below 18 years of age.

[3] https://censusindia.gov.in/2011-prov-results/data_files/india/Final_PPT_2011_chapter5.pdf (accessed on April 20, 2021)

[4] https://censusindia.gov.in/Data_Products/Library/Provisional_Population_Total_link/PDF_Links/chapter6.pdf (accessed on April 20, 2021)

[5] This information can be retrieved from http://rchiips.org/nfhs/data/hr/hrfctsum.pdf (accessed on April 21, 2021)

[6] Refer to Reserve Bank of India (RBI) website: https://www.rbi.org.in/scripts/PublicationsView.aspx?id=17923 (accessed on April 21, 2021)

[7] This is based on average exchange rate for 2015-16 obtained from RBI.

[8] While estimating the program impact, we also used other indicators of poverty apart from BPL card possession. The findings are robust to these alternate criteria of identification of the poor (Section 5.2.4).

[9] MPCE is considered to be an accepted indicator of income and it has been used to estimate national and state-wise poverty rates in India (Deaton, 2003; Meyer and Sullivan, 2012)

[10] We used other cut-offs as well but our results as found to be robust (Section 5.3)

[11] The details of all the variables used in the regressions are given in table A1 of the appendix.

[12] For more information, refer to https://unstats.un.org/unsd/demographic-social/time-use/icatus-2016/ (accessed on April 17, 2021)

[13] State specific interventions may include and not limited to school construction, improving school infrastructure or general increase in demand for education owing to increase in household income among many others.

[14] One such example might be reduced crime against women which can fetch higher gains in years of schooling for females in comparison to the males in the same age cohort.

[15] Muralidharan and Prakash (2017) uses similar argument to use Jharkhand as a control state to Bihar

[16] Refer https://censusindia.gov.in/2011-prov-results/data_files/india/Final_PPT_2011_chapter6.pdf l (accessed on April 20, 2021)

[17] Punjab and Haryana together contribute over 67% of wheat in the country. Refer to https://thewire.in/agriculture/punjab-protests-rice-wheat-procurement-ten-year-plan-foood-security (Accessed on April 20, 2021)

[18] In the section on robustness checks (section 5.3), we use different combination of districts from states neighboring Haryana instead of Punjab and find that the findings are robust.

[19] For more information on the survey, refer to http://rchiips.org/nfhs/nfhs3.shtml (accessed on April 17, 2021).

[20] Haryana districts: Ambala, Fatehbad, Jind, Kaithal, Kurukshetra, Panchkula, Sirsa; Punjab districts: Bhatinda, Mansa, Mukhtsar, Patiala, Sangrur, SAS Nagar

[21] Haryana districts: Ambala, Fatehbad, Jind, Kaithal, Kurukshetra, Panchkula, Sirsa, Yamunanagar, Punjab districts: Bhatinda, Mansa, Mukhtsar, Patiala, Sangrur, SAS Nagar; Himachal Pradesh districts: Sirmaur, Solan; Uttarakhand district: Dehradun

[22] Haryana districts: Ambala, Faridabad, Fatehbad, Jind, Kaithal, Kurukshetra, Mewat, Palwal, Panchkula, Panipat, Sirsa, Sonipat, Yamunanagar; Punjab districts: Bhatinda, Mansa, Mukhtsar, Patiala, Sangrur, SAS Nagar
Himachal Pradesh districts: Sirmaur, Solan; Uttarakhand district: Dehradun; Uttar Pradesh districts: Aligarh, Baghpat, Gautam BudhaNagar, Ghaziabad, Mathura, Muzzafarnagar, Saharanpur.

[23] The parallel trends regression output can be provided on request.

[24] The regression to test parallel trend assumption can be provided on request.

[25] These findings can also be replicated by using PLFS as well. The results can be provided on request.

[26] The regression results where we test the parallel trends assumption are available on request.

[27] The results from the regressions where the parallel trends assumption are tested are available on request.

[28] Regression results for test parallel trends assumption are available on request.

[29] As indicated earlier, the beneficiaries can receive Rs. 30,000 or Rs, 35,000 in case the maturity is deferred by two years or four years respectively, which implies they remain unmarried till 20 or 22 years of age.